\documentclass[pra,onecolumn,amsmath,amssymb,floatfix,nofootinbib,secnumarabic,showpacs]{revtex4}

\usepackage{graphics}      
\usepackage{graphicx}      
\usepackage{longtable}     
\usepackage{url}           
\usepackage{bm}            
\usepackage{bbm}
\usepackage[dvips]{epsfig,color}
\usepackage{dcolumn}

\def \c{\hat{c}}
\def \a{\hat{a}}

\def \b{\hat{b}}
\def\la{\langle}
\def\ra{\rangle}
\def \beq{\begin{equation}}
\def \eeq{\end{equation}}
\def \beqarr{\begin{eqnarray}}
\def \eeqarr{\end{eqnarray}}
\def \bspt{\begin{split}}
\def \espt{\end{split}}
\def \bef{\begin{figure}}
\def \enf{\end{figure}}

\newcommand{\abs}[1]{\lvert#1\rvert}

\newcommand{\ev}[1]{\mbox{$\langle #1 \rangle$}}

\newcommand{\ket}[1]{\mbox{$| #1 \rangle$}}

\makeatletter
\renewcommand*\env@cases[1][1.2]{%
  \let\@ifnextchar\new@ifnextchar
  \left\lbrace
  \def\arraystretch{#1}%
  \array{@{}l@{\quad}l@{}}%
}
\makeatother

\begin{document}

\title{Entanglement Entropy and Mutual Information in Bose-Einstein Condensates}

\author{Wenxin Ding}
\affiliation{NHMFL and Department of Physics, Florida State
University, Tallahassee, Florida 32306, USA}

\author{Kun Yang}
\affiliation{NHMFL and Department of Physics, Florida State
University, Tallahassee, Florida 32306, USA}

\date{\today}

\begin{abstract}
In this paper we study the entanglement properties of free {\em
  non-relativistic} Bose gases. At zero temperature, we calculate the bipartite block
entanglement entropy of the system, and find that it diverges
logarithmically with the particle number in the subsystem. For
finite temperatures, we study the mutual information between the two
blocks. We first analytically study an infinite-range hopping model,
then numerically study a set of long-range hopping models in
one-dimension that exhibit Bose-Einstein condensation. In both cases
we find that a Bose-Einstein condensate, if present, makes a
divergent contribution to the mutual information which is
proportional to the logarithm of the number of particles in the
condensate in the subsystem. The prefactor of the logarithmic
divergent term is model dependent.
\end{abstract}


\maketitle


\section{Introduction}

Entanglement, as measured by, {\em e.g.}, bipartite block entanglement entropy, is
playing an increasingly important role in the study of
condensed-matter or quantum many-body physics, both conceptually and
quantitatively. It has been  used as a very useful and in
some cases indispensable way to characterize phases and phase
transitions, especially for phases and quantum phase transitions in
strongly correlated fermionic or spin systems (for a review, see Ref. [\onlinecite{amico:517}]).
For bosonic systems, studies of entanglement entropy have mostly
focused on {\em relativistic} free bosonic field theories \cite{wilczek94,calabrese-2004-0406,casini-2005-0512},
which are equivalent to coupled harmonic oscillator systems (for reviews,
see Refs. [\onlinecite{adesso-2007-40}] and [\onlinecite{RevModPhys.77.513}]).

In this paper we study the entanglement properties of free {\em
  non-relativistic} Bose gases. In addition to interest in its own
right, our motivation also comes in part from the
following consideration. In recent studies it has been shown that
entanglement is enhanced at quantum critical points \cite{QPT_Sachdev} and
strongly correlated phases with topological order \cite{xgwen1990}, in the form of
either violation of area law \cite{PhysRevD.34.373, area-law,
  eisert-2008, wolf:010404, gioev:100503,  Barthel:2006, Li:2006, wilczek94,
  PhysRevLett.90.227902, calabrese-2004-0406, PhysRevLett.93.260602,
  santachiara-2006-L06002, feiguin:160409, bonesteel:140405},
or subleading corrections to the area law that diverges with block
size \cite{levin:110405, kitaev:110404, fradkin:050404} (usually in a
logarithmic fashion). On the other hand there have been relatively
few studies of the behavior of entanglement entropy in states with
traditional long-range order \cite{vidal05, vidal06, vidal07}. In a
recent work \cite{ding:052109}, we calculated the block entanglement
entropy of some exactly soluble spin models that exhibit
ferromagnetic or antiferromagnetic long-range order in the ground
state, and found that such conventional orders also lead to
logarithmically divergent contribution to the entropy. Bose-Einstein
condensation (BEC) is perhaps the simplest example of conventional
ordering. It is thus natural to study its entanglement properties.
As we are going to show, a Bose-Einstein condensate (referred to as
a condensate from now on) indeed makes a logarithmically divergent
contribution to the entropy as well.

Besides the entanglement entropy of the ground state, the entanglement
properties of system at finite temperature are also of great interest. However, the entanglement
entropy is only well-defined for a pure state. For a system that is
described by a mixed density matrix, the von Neumann entropy of the
reduced density matrix becomes different for the two parts of the
bipartite systems. In such cases, there is a natural extension of the
entanglement entropy that one can work with - the mutual information
\cite{dreissig,Wolf2008}. We will show that a condensate, when
present, makes a logarithmically divergent contribution to the mutual
information.

This paper is organized as follows. In Sec. II we study the ground state
entanglement entropy of a generic free boson model that is
translationally invariant \cite{klich06}. In Sec. III we introduce an infinite-range
hopping model for bosons which is exactly solvable, and calculate the
mutual information analytically. In Sec. IV we introduce a long-range
hopping model for bosons in one-dimension (1D) which exhibits a finite temperature BEC
for a certain parameter range, then we present a numerical
study of the mutual information for this model. In the end, we
summarize and discuss the results of this paper in Sec. V.

\section{Zero Temperature: Entanglement Entropy of Free Bosons}
Consider a general Hamiltonian of free bosons hopping on a lattice
of size $L$: \beq\label{Eq:GeneralH} H = - \sum_{ij} t_{ij}
\a^\dagger_i \a_j, \eeq where $t_{ij} > 0$, $\a_i(\a^\dagger_i)$'s
are the bosonic annihilation (creation) operators. If the system is
translationally invariant, $t_{ij} = t_{i-j}$, then the Hamiltonian
can be diagonalized by Fourier transformation:
\beq H = \sum_{k}
\varepsilon(k) \b^\dagger_k \b_k, \eeq 
where $\b_k = \frac{1}{\sqrt{L}}\sum_j e^{-ijk} \a_j$ is the annihilation operator
in $k$ space. In most generic cases, the ground state is the $k=0$
state. At zero temperature, all particles fall into the ground
state. For a system containing $N$ particles, the ground state is
given by:
\beq \ket{\Psi_0} = \frac{1}{\sqrt{N!}} (\b^\dagger_0)^N
\ket{0} = \frac{1}{\sqrt{N!}} (\frac{1}{\sqrt{L}} \sum_j
\a^\dagger_j)^N \ket{0}. \eeq

To consider its bipartite block entanglement entropy, we divide the
system of size $L$ in two parts, and label them $A$ and $B$
respectively. Let the
sizes of each part be $L_A$ and $L_B$, $L_A + L_B = L$, and define
\beq
\a^\dagger_A = \frac{1}{\sqrt{L_A}} \sum_{j \in A}
\a^\dagger_j,\ \a^\dagger_B = \frac{1}{\sqrt{L_B}} \sum_{j \in B}
\a^\dagger_j.
\eeq
Then we can write $\ket{\Psi_0}$ as:
\beq
\begin{split}
\ket{\Psi_0} & = \frac{L^{-N/2}}{\sqrt{N!}} (\sqrt{L_A} \a_A +
\sqrt{L_B} \a_B )^N \ket{0} = \frac{L^{-N/2}}{\sqrt{N!}}\sum_{l=0}^{N}
\frac{N!}{(N-l)!l!} (\sqrt{L_A} \a^\dagger_A)^{l} (\sqrt{L_B}
\a^\dagger_B)^{N-l} \\
& = L^{-N/2} \sum_{l} \sqrt{\frac{N!}{(N-l)!l!}} L_A^{l/2}
L_B^{(N-l)/2}  \left[ \frac{1}{\sqrt{l!}} \a^{\dagger l}_A \frac{1}{\sqrt{(N-l)!}}
\a^{\dagger N-l}_B \ket{0} \right] \\
& = \sum_l \sqrt{\lambda_l} \ket{l}_A \otimes \ket{N-l}_B,
\end{split}
\eeq
where $\lambda_l = L^{-N} \frac{N!}{(N-l)!l!} L_A^{l} L_B^{N-l}$,
$\ket{l}_A = \frac{1}{\sqrt{l!}} \a^{\dagger l}_A \ket{0}_A$,
$\ket{N-l}_B = \frac{1}{\sqrt{(N-l)!}} \a_B^{^\dagger N-l}\ket{0}_B$,
and $\ket{0} = \ket{0}_A \otimes \ket{0}_B$.

This is an explicit Schmidt decomposition, and therefore the
entanglement entropy is readily given by:

\beq
E = - \sum_l \lambda_l \ln \lambda_l.
\eeq

We are interested in the asymptotic behavior in two limiting cases: (1) the
equal partition case; (2) size of $B$ is substantially larger than
$A$, i.e., $L_B \gg L_A$.

(i) Equal partition, $L_A = L_B = \frac{L}{2}$:
\beq
\lambda_l = \frac{N!}{l!(N-l)!} \frac{L_A^l L_B^{N-l}}{L^N} =
\frac{N!}{l!(N-l)! 2^N} = \frac{N!}{(\frac{N}{2})!}
\frac{(\frac{N}{2})! 2^N}{(\frac{N}{2} - (\frac{N}{2} - l))! (\frac{N}{2}
  + (\frac{N}{2} - l))!}.
\eeq
Let $x = \frac{N}{2} - l$, then $x \in [-\frac{N}{2},
  \frac{N}{2}]$, and we can denote $\lambda_l$ as $\lambda_x =
\frac{2^N N!}{(\frac{N}{2})!} \frac{(\frac{N}{2})!}{(\frac{N}{2} -
  x)!(\frac{N}{2} + x)!}$ which can be approximated by a Gaussian
distribution factor
$\lambda_x \sim e^{\frac{- 2x^2}{N}}$ when $N$ is large.
In the limit $N \rightarrow \infty$, the summation
over $n$ (or $x$) can be approximated by an integral. Also in this
limit, the Gaussian factor is sharply peaked around $x = 0$, the
integral region can be extended to from minus infinity to
infinity. Using the fact that $\sum_x \lambda_x\simeq
\int^{\infty}_{-\infty}\lambda(x) dx = 1$, we arrive at
\beq
\lambda(x) \simeq \sqrt{\frac{2}{N \pi}} e^{-\frac{2 x^2}{N}}.
\eeq

The entanglement entropy is then
\beq \label{eqn:EEEP}
E \simeq - \int^{\infty}_{-\infty}\lambda(x)\ln \lambda(x) dx =
\frac{1}{2} \left(1 + \ln(\frac{N \pi}{2}) \right) = \frac{1}{2} \ln N
+ \mathcal{O}(1).
\eeq

(ii) Unequal partition, $L_B \gg L_A$:

If $L_B \gg L_A$, $L \rightarrow \infty$, but keep $\frac{N}{L}
\rightarrow \ev{n}$ (fixed), the distribution of $\lambda_l$
approaches a Poisson distribution:
\beq
\lambda_l = \frac{N!}{l!(N-l)!} \frac{L_A^l L_B^{N-l}}{L^N}
\xrightarrow{^{N \rightarrow \infty}} \frac{(L_A \ev{n} )^l e^{-L_A \langle n \rangle }}{l!}.
\eeq
The entropy of the Poisson distribution, which in this case is our
entanglement entropy, is known to be:
\beq\label{Eq:EE}
\begin{split}
E &= \frac{1}{2} [1 + \ln (2\pi L_A \ev{n})] - \frac{1}{12 L_A \ev{n}} +
O(\frac{1}{(L_A \ev{n})^2})\\
&= \frac{1}{2}\left[1 + \ln (2 \pi N_A )\right] -\frac{1}{12 N_A} +
  O(\frac{1}{(N_A)^2}) = \frac{1}{2}\ln N_A + \mathcal{O}(1),
\end{split}
\eeq
where $N_A = L_A \ev{n}$ is the average particle number in subsystem $A$.

Therefore, we find, in both cases, that the leading term of the
mutual information goes as $\frac{1}{2} \ln N_A$ for $L_A \le L_B$.

%
%

\section{Mutual Information: Analytic Study of an Infinite-Range Hopping Model}
In this section, we will study the natural generalization of
entanglement entropy at finite temperature: the mutual
information, which is defined as
\beq E_M = \frac{1}{2} (E_A + E_B - S), \eeq
where $E_A$ and $E_B$ are the von Neumann entropy of the reduced density
matrices of subsystems A and B, respectively, and $S$ is the entropy of
the whole system. Note that at finite temperature $E_A$ and $E_B$ are no
longer the same due to the fact that the system is described by a mixed
density matrix. We must emphasize here that our definition of
mutual information differs from its usual definition \cite{minote} by a factor of 2
so that it will converge to the entanglement entropy when the system
approaches a pure state.

\subsection{Model, spectrum and thermodynamic properties}
In order to facilitate an exact solution, we consider the following
infinite-range hopping model which is obtained by setting $t_{ij}$
in Eq. (\ref{Eq:GeneralH}) to a constant properly scaled by the
system size $t_{ij} = t / L$ so that the thermodynamic limit is
well-defined. The Hamiltonian is then

\beq
H = -\frac{t}{L} \sum_{i,j}\a_i^\dagger \a_j = - \frac{t}{L} (\sum_i
\a_i^\dagger) (\sum_j \a_j).
\eeq
By substituting the Fourier transform of $\a_j$'s defined in Sec. II,
$\b_k = \frac{1}{\sqrt{L}}\sum_j e^{-ijk} \a_j$, one obtains:
\beq
H = -t \b^\dagger_0 \b_0.
\eeq
This model has a very simple spectrum with a ground state with energy
$-t$, and all the other excited states are degenerate with zero
energy. This particularly simplified spectrum makes an exact solution possible.

To study the finite temperature properties of this model, we will work
with the grand canonical ensemble (GCE), 
in which the chemical potential            
$\mu$ is introduced to control the average 
density of the system. This model exhibits BEC
at finite temperature $T_C$. To determine $T_C$, we start by
considering a system of finite size $L$, and its occupation numbers are:
\beq
\ev{N_{k=0}} = \ev{N_0} = \frac{1}{e^{\beta (-t -
    \mu)} - 1},\text{ } \ev{N_{k}} = \frac{1}{e^{\beta(-\mu)} -
  1}\text{ for $k \ne 0$}.
\eeq
Here $\ev{N_{0}}$ and $\ev{N_{k}}$ denote average occupation numbers for
the corresponding states in $k$-space; $\beta = \frac{1}{T}$ is the
inverse temperature. From this point on, when we write
$\ev{N_k}$, it immediately indicates $k \ne 0$. The average total particle number of the
system will be denoted as $\ev{N}$. To identify $T_C$, we know in the
thermodynamic limit, when $T \rightarrow T_C + 0^+$, $\mu \rightarrow
E_{k=0} = -t$, and $\frac{\la N_0 \ra}{N} \rightarrow 0$. Therefore,
$\ev{N_{k}} = \frac{1}{e^{\beta_c t} - 1} \simeq \frac{ \la N \ra }{L}
= \ev{n}$ where $\ev{n}$ is the average particle density. So we
obtain:
\beq
T_C = \frac{t}{\ln (1 + 1/\ev{n})}.
\label{Eq:inf_Tc}
\eeq

Above $T_C$, $\ev{n} = \frac{L - 1}{L} \frac{1}{e^{-\beta \mu} - 1} +
\frac{1}{L} \frac{1}{e^{\beta (-t - \mu)} - 1}$, from which in the
large $L$ limit we can derive that
\beq
\mu = -T \ln\left(1+\frac{1}{\ev{n}}\right).
\eeq
$\mu$ has a finite size correction which is negligible above $T_C$,
but will become important below $T_C$.

We also know that the partition function of the system in GCE. bears
the following form:
\beq
Z = (\frac{1}{1 - e^{\beta \mu}})^{L-1} \frac{1}{1 - e^{\beta(t+\mu)}},
\eeq
from which it is easy to show that the entropy in GCE. takes the following form:
\beq\label{Eq:thermal entropy}
\begin{split}
S &= - \frac{\partial \Omega}{\partial T} = \ln Z - \frac{1}{T}
\frac{\partial}{\partial \beta} \ln Z\\
&= (1 + \ev{N_0}) \ln(1 + \ev{N_0}) - \ev{N_0} \ln N_0 + (L - 1)
\left[(1 + \ev{N_k}) \ln (1 + \ev{N_k}) - \ev{N_k} \ln \ev{N_k} \right].
\end{split}
\eeq

Anticipating later relevance, we are particularly interested in the behavior of
finite size systems near $T_C$. For a finite system, the chemical
potential $\mu$ is no longer strictly equal the ground state energy
below $T_C$, but picks up a finite size correction $\delta \mu$
determined by the following condition:
\beq
\frac{1}{e^{-\beta \delta \mu} - 1} = \ev{N_0},
\eeq
from which we can easily derive that
\beq
\delta \mu = - T \ln\left(1 + \frac{1}{\ev{N_0}}\right).
\eeq
Consider $T = T_C = \frac{t}{\ln(1 + 1/\la n \ra)}$, and make use of the
following fact
\beq
\ev{N_0}= \ev{N} - \sum_{k \ne 0} \ev{N_k} = \ev{N} - \frac{L -
  1}{e^{\beta (t - \delta \mu)} - 1},
\eeq
we obtain
\beq
\ev{N_0} = \ev{N} - \frac{L - 1}{e^{\beta_c (t - \delta \mu)}
  - 1} = \ev{N} - \frac{L - 1}{(1 + L / \ev{N}) (1 + 1 / \ev{N_0}) - 1}.
\eeq
This equation can be solved to give $\ev{N_0}$ as a function of system
size $L$ at a given density $\ev{n} = \ev{N} / L$, at $T=T_C$:
\beq\label{Eq:N_0@Tc}
\ev{N_0} = \sqrt{L} \sqrt{\left(\frac{\ev{N}}{L}\right)^2 +
  \frac{\ev{N}}{L} + \frac{1}{4L}} \simeq \sqrt{L} \sqrt{ \langle n
  \rangle ^2 + \langle n \rangle }.
\eeq
Even though this divergent $N_0$ does not affect the thermodynamic behavior
of the system, as we will see later it makes a (leading) divergent contribution to the mutual information
at $T=T_C$ depending on how the system is partitioned, or specifically
how large is the subsystem size $L_A$ compared with this $\sqrt{L}$ divergence.

\subsection{Formalism and issues}
In the following part, we will use Peschel's result \cite{Peschel2003}
on the reduced density matrix of a Gaussian state:
\beq\label{Eq:RDM}
\rho_A = \mathcal{K} e^{\{\ln((1+G)G^{-1})\}^T_{ij}\a^\dagger_i \a_j},
\eeq
where $G_{ij} = \ev{\a^\dagger_i \a_j}$ is the two point correlation
function matrix truncated within the subsystem, and $\mathcal{K}$ is the
normalization factor. The entropy is given as
\beq\label{Eq:entropy}
E_A = \sum_l \left[(1 + g_l)\ln (1 + g_l) - g_l \ln g_l\right],
\eeq
where $g_l$'s are the complete set of eigenvalues of $G$'s (after
truncation). Actually this formula also applies to the original
system.

We must note that, this formula does not lead to the correct
zero temperature limit for the entropy. At zero temperature,
$G_{ij} = \ev{a_i^\dagger a_j} = \ev{n}$. Its eigenvalues are all zero
except for one: $g_0 = \ev{n} L = \ev{N}$, which gives us a non-zero entropy
$S_{T=0} =(\ev{N} + 1) \ln (\ev{N} + 1) - \ev{N} \ln \ev{N} = \ln \ev{N} + \ev{N} \ln (1 +
\frac{1}{\ev{N}}) \sim \ln \ev{N}$ at $T = 0$. This reflects
the fact that we are working with GCE where the particle-number
fluctuation is still permissible at $T = 0$ and the fluctuation
amplitude $\delta N \sim \ev{N}$. However, as we show below, the mutual information still converges to
the correct zero temperature limit, the entanglement entropy, at least
to the leading order.

The von Neumann entropy for a subsystem $A$ is given by
\beq
E_{A}^{(\text{GCE})} = (N_A + 1) \ln (N_A + 1) - N_A \ln N_A = \ln N_A + N_A \ln (1 +
\frac{1}{N_A}),
\eeq
where $N_A = \ev{n} L_A$ is the average total particle number in the subsystem
$A$. In the large $N$ limit, the second term converges to $1$. So the
mutual information is given by:
\beq
E_{M} \equiv \frac{1}{2} (E_A + E_B - S_{GCE\ T=0}) = \frac{1}{2} \left(\ln
\frac{N_A N_B}{N} + 1\right).
\eeq
For $N_A \le N_B$, we have
\beq
E_M \simeq \frac{1}{2} \ln N_A + \mathcal{O}(1).
\eeq
This agrees with Eq. (\ref{Eq:EE}) at the leading order.

\subsection{Mutual information}

According to our Eq. (\ref{Eq:RDM}) and Eq. (\ref{Eq:entropy}), to
obtain the von Neumann entropy of the reduced density matrix, all what we
have to do is to diagonalize the truncated two-point correlation
function matrix. Fortunately, within this infinite-range hopping
model, this is rather simple. For a finite system, we can obtain a
general result for all temperatures:
\beq
\begin{split}
G_{ij} &= \ev{\a^\dagger_i \a_j} = \ev{\frac{1}{L} \sum_k e^{-i k (i-j)}
  \b^\dagger_k \b_k} \\
& = \frac{1}{L} \ev{\b^\dagger_0 \b_0} + \frac{1}{L} \sum_{k \neq 0}
e^{-ik(i-j)}\ev{\b^\dagger_k \b_k} = \frac{\ev{N_{0}}}{L} +
\frac{\ev{N_{k}}}{L} \sum_{k \neq 0} e^{-ik(i-j)}\\
& = \frac{\ev{N_0}}{L} + (\delta_{ij} - \frac{1}{L}) \ev{N_k}.
\end{split}
\eeq
In the above calculation, we have made use of the fact that $\ev{N_k}$
is $k$-{\em independent}. This matrix is easily diagonalized. For a
system of size $L$, and a $G$ truncated to a size of $L_A \times L_A$
denoted by $G_{A}$, the eigenvalues are
\beq
g_1 = \frac{L_A \ev{N_0}}{L} + \frac{L - L_A}{L} \ev{N_k},\ g_l =\ev{N_k} \text{
  for $l = 2, \dots , L_A$}.
\eeq
Now the von Neumann entropy of subsystem $A$ can be calculated
directly from above result:
\beq
\begin{split}
E_A &= \sum_{l = 1}^{L_A} \left((1 + g_l) \ln (1 + g_l) - g_l \ln g_l
\right)\\
&= \left( 1 + \frac{L_A \ev{N_0}}{L} + \frac{L - L_A}{L} \ev{N_k} \right) \ln
\left(1 + \frac{L_A \ev{N_0}}{L} + \frac{L - L_A}{L} \ev{N_k} \right) \\
&- \left(\frac{L_A \ev{N_0}}{L} + \frac{L - L_A}{L} \ev{N_k} \right)
\ln \left(\frac{L_A \ev{N_0}}{L} + \frac{L - L_A}{L} \ev{N_k} \right)\\
& + (L_A - 1)\left[(1 + \ev{N_k}) \ln (1 + \ev{N_k}) - \ev{N_k} \ln
(\ev{N_k}) \right].\\
\end{split}
\eeq
Combining the above with Eq. (\ref{Eq:thermal entropy}), we can obtain the mutual
information for a general bipartite system:
\begin{equation}\label{Eq:MI}
\begin{split}
&E_M = \frac{1}{2} (E_A + E_B - S) \\
&= \frac{1}{2} \biggl[ \left(1 + \frac{L_A
  \ev{N_0}}{L} + \frac{L_B}{L} \ev{N_k}\right) \ln\left(1 + \frac{L_A
  \ev{N_0}}{L} + \frac{L_B}{L} \ev{N_k}\right)\\
& - \left(\frac{L_A \ev{N_0}}{L} + \frac{L_B}{L} \ev{N_k}\right) \ln\left(\frac{L_A
  \ev{N_0}}{L} + \frac{L_B}{L} \ev{N_k}\right) \\
&+ \left(1 + \frac{L_B \ev{N_0}}{L} + \frac{L_A}{L} \ev{N_k}\right) \ln\left(1 + \frac{L_B
  \ev{N_0}}{L} + \frac{L_A}{L} \ev{N_k}\right)\\
&-\left(\frac{L_B \ev{N_0}}{L} + \frac{L_A}{L} \ev{N_k}\right) \ln\left(\frac{L_B
  \ev{N_0}}{L} + \frac{L_A}{L} \ev{N_k}\right) \\
&  -(1 + \ev{N_k}) \ln (1 + \ev{N_k}) + \ev{N_k} \ln \ev{N_k} - (1 +
\ev{N_0}) \ln (1 + \ev{N_0}) + \ev{N_0} \ln \ev{N_0} \biggl]. \\
\end{split}
\end{equation}

Next, we shall discuss the asymptotic behavior of $E_M$ in different
temperature regions and with different partitions.

\begin{enumerate}
\item[(1)] $L_A \ll L$, $T > T_C$:

in this case, $\ev{N_0}$ and $\ev{N_k} \simeq \ev{n}$ are both of
order one , so $\frac{L_A \la N_0 \ra }{L} \rightarrow 0$, $\frac{L_A
  \la N_k \ra }{L} \rightarrow 0$, $L_B \simeq L$. Thus
\beq
E_M \simeq \mathcal{O}\left(\frac{L_A}{L}\right).
\eeq

\item[(2)] $L_A = L_B = \frac{L}{2}$, $T > T_C$:
in this case, $E_M$ is reduced to
\beq
\begin{split}
E_M &= \frac{1}{2} \biggl(\left(2 + \ev{N_0} + \ev{N_k}\right) \ln \left(1 +
\frac{1}{2}\left(\ev{N_0} + \ev{N_k}\right)\right) \\
&- \left(\ev{N_0} + \ev{N_k}\right) \ln \left(\ev{N_0}/2 + \ev{N_k}/2\right) \\
& - (1 + \ev{N_k}) \ln\left(1 + \ev{N_k}\right) +\ev{N_k} \ln \ev{N_k} \\ 
&- \left(1 + \ev{N_0}\right) \ln (1 + \ev{N_0}) + \ev{N_0} \ln \ev{N_0} \biggl)
\end{split}
\eeq
which can also be written as an explicit function of $L$, $\ev{n}$ and $T$ using previous results.
\item[(3)] $L_A \ll L$, $T < T_C$:
in this case, $\ev{N_0}$ becomes a macroscopic number, while $\ev{N_k}$ remains
 to be of order $1$. Therefore we isolate the contribution from \ev{N_0},
and other terms are of order $\mathcal{O}(1)$:
\beq\label{Eq:E_m3}
E_M = \frac{1}{2} \ln \left(1 + \frac{L_A \la N_0 \ra }{L}
\right) + \mathcal{O}(1).
\eeq
If we define $\ev{N_{A0}} = \frac{L_A \la N_0 \ra }{L}$ to be the
average particle number in the condensate of the subsystem, then
\beq
E_M = \frac{1}{2} \ln \ev{N_{A0}} + \mathcal{O}(1).
\eeq
\item[(4)] $L_A = L_B = \frac{L}{2}$, $T < T_C$:
the leading contribution is again obtained by keeping $\ev{N_0}$'s contribution only:
\beq\label{Eq:E_m4}
E_M = \frac{1}{2} \ln \left(\frac{1}{4}\ev{N_0} + 1 \right) + \mathcal{O}(1)
= \frac{1}{2} \ln \ev{N_{A0}} + \mathcal{O}(1).
\eeq

\item[(5)] $L_A \ll L$, $T = T_C$:
as we calculated before, \ev{N_0} diverges as $\sqrt{L}$. When $L_A
\ll L$, according to Eqs. (\ref{Eq:E_m3}) and (\ref{Eq:N_0@Tc}),
we have
\beq
E_M = \frac{1}{2} \ln \left(1 + \frac{L_A \sqrt{(\la n
\ra + 1) \la n \ra}}{\sqrt{L}} \right) + \mathcal{O}(1) = \frac{1}{2}
\ln\left(\la n \ra \frac{L_A}{\sqrt{L}}\right) + \mathcal{O}(1).
\eeq
For such partition, the scaling behavior of mutual information depends
on the ratio $\frac{L_A}{\sqrt{L}}$. If we consider $L_A$ is a small
but still finite fraction of $L$, the scaling behavior of the mutual
information still persists: $E_M = \frac{1}{4} \ln (\la n \ra L_A) + \mathcal{O}(1)$.

\item[(6)] $L_A = L/2$, $T = T_C$:
by referring to Eqs. (\ref{Eq:E_m4}) and (\ref{Eq:N_0@Tc}), we
have \beq E_M = \frac{1}{2} \ln \left(\frac{1}{4} \sqrt{(\la n \ra +
1)\la
  n \ra L} \right) + \mathcal{O}(1) = \frac{1}{4} \ln (\la n \ra L_A) + \mathcal{O}(1).
\eeq
\end{enumerate}

To sum up, from this study, we find that the extensive part of the
thermal entropy of the whole system is canceled out in the mutual
information. Below $T_C$ the mutual information is dominated by
contribution from the condensate. Even above $T_C$, contribution
from the condensate is of the same order as that from the excited
states. The mutual information really characterizes the quantum
feature of the system.

\begin{figure}
  \includegraphics[width=8cm]{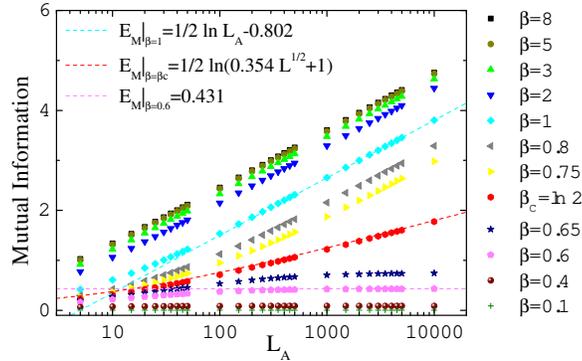}
  \caption{(Color online) Numerical calculation of mutual information
    for the infinite-range hopping model for equally partitioned
    systems with average density $\ev{n} = 1$. The scatters are
    numerical data, while the dash lines are obtained from our analytic
    results corresponding to the particular temperature. We see
    exactly what our analytic results tell us: above $T_C =
    \frac{1}{\ln 2}$ the mutual information saturates; below
    $T_C$, $E_M \simeq \frac{1}{2} L_A$; at $T_C$, $E_M \simeq
    \frac{1}{4} L_A$. Note that the analytic results (dash lines) deviate
    from the numerical calculation because we only keep terms to the
    subleading order; terms that goes to zero [i.e., of order
    $\mathcal{O}(\frac{1}{L_A})$] in the thermodynamic limit are
    neglected.}
  \label{Fig:mi_inf}
\end{figure}

To visualize the behavior of mutual information, we present a numerical calculation for the
mutual information of this model in Fig. (\ref{Fig:mi_inf}). This is
done by numerically diagonalizing the truncated two-point correlation function
matrix then computing the von Neumann entropy of the reduced density
matrix from those eigenvalues. The system is equally partitioned,
$\ev{n} = 1$, so $T_C = \frac{1}{\ln 2}$. We see exactly what our analytic
results tell us: above $T_C$ the mutual information saturates; below
$T_C$, $E_M \simeq \frac{1}{2} L_A$; at $T_C$, $E_M \simeq \frac{1}{4}
L_A$. Note that in the plot, the analytic results (dash lines) deviate
from the numerical calculation because we only keep terms to the
subleading order; terms that goes to zero [i.e. of order
$\mathcal{O}(\frac{1}{L_A})$] in the thermodynamic limit are
neglected.

\section{Numerical Study of Mutual Information in One-dimension with Long-range Hopping}

In this section, we shall present our results of numerical study of
the mutual information of free bosons living on a one dimensional
lattice. In this numerical study, we adopt our previous method, and
calculate the von Neumann entropy of the reduced density matrix from
the eigenvalues of the truncated two-point correlation function
matrix. Throughout this calculation, we hold the average density
fixed at $\ev{n} = 1$ (which means we keep adjusting the chemical potential at
different temperatures) and consider equal partition only.

It is well known that for nearest-neighbor (NN) (or other short-range) hopping models whose
dispersion relation at long-wave length takes the form $\epsilon (k) \sim k^2$, a finite
temperature BEC can only exist in three dimensions (3D). However, 3D is in general very
challenging for a numerical study that requires large system sizes.
Moreover, in 3D the mutual information is dominated by area
law \cite{area-law}, which renders the logarithmic divergence
suggested by our study in Sec. III sub-leading and thus difficult to
isolate. For both of these reasons, it is
desirable to study a model in 1D with BEC at finite $T$. In 1D, the
short-range hopping model does {\em not} support BEC at finite
temperature. To stabilize a condensate in 1D, we introduce power-law
long-range hopping in our free boson model to modify its long-wave
length dispersion. This is similar to what was done in
Ref. [\onlinecite{yusuf}], in which the authors introduced long-range
interaction between spins to stabilize magnetic order in 1D. The
Hamiltonian with long-range hopping is obtained by setting $t_{ij}$ in
Eq. (\ref{Eq:GeneralH}) to the following form, tuned
by a parameter $\gamma$:
\beq
H = -\sum_{ij} \frac{t}{\abs{i-j}^\gamma} a^\dagger_i a_j = -2t \sum_k
(\sum_{n=1}^{L-1} \frac{\cos (nk)}{n^\gamma}) b^\dagger_k b_k = \sum_k
\varepsilon_\gamma(k) b^\dagger_k b_k.
\eeq
We will show that the long wave-length dispersion is modified to be
$\varepsilon_\gamma(k) \sim k^{\gamma - 1}$  for $\gamma < 3$, as a
result if which a finite temperature BEC exists for $\gamma < 2$.

Consider the eigenenergy function $\varepsilon_\gamma(k) =
- 2 t \sum_{n=1}^{L-1} \frac{\cos (nk)}{n^\gamma}$ in the thermodynamic
limit:
\beq
\begin{split}
\varepsilon_\gamma(k) = -2t\ \sum_{n=1}^{\infty} \frac{\cos (nk)}{n^\gamma}
= -2t\ \text{Re}\left[ \sum_n \frac{e^{i n k}}{n^\gamma} \right]
= -2t\ \text{Re}\left[ F(\gamma, i k) \right],
\end{split}\label{Eq:ek}
\eeq
where $F(\gamma, v)$ is the Bose-Einstein integral function \cite{PhysRev.83.678} defined as:
\beq
F(\gamma, v) = \frac{1}{\Gamma(\gamma)} \int dx \frac{x^{\gamma -
    1}}{e^{x + v} - 1} = \sum_{n=1}^{\infty} \frac{e^{- n v}}{n^\gamma}.
\eeq
The analytic properties of $F(\gamma, v)$ near $v = 0$ are known \cite{PhysRev.83.678}:
\begin{equation}\label{Eq:vare_expansion}
F(\gamma, v) =
\begin{cases}[1.5]
  \displaystyle{\Gamma(1 - \gamma) v^{\gamma - 1} + \sum_{n=0}^{\infty}
  \frac{\zeta(\gamma - n)}{n!} (-v)^n}, &
  \gamma \notin \mathbb{Z}, \\
  \displaystyle{\frac{(-v)^{\gamma - 1}}{(\gamma - 1)!} \left[\sum_{r = 1}^{\gamma -
      1} \frac{1}{r} - \ln(v) \right] + \sum_{n \neq \gamma -
    1}^{\infty} \frac{\zeta(\gamma - n)}{n!} (-v)^n}, &
  \gamma \in \mathbb{Z},
\end{cases}
\end{equation}
where $\zeta(x)$ is the Riemann zeta function. Thus we find that
$\varepsilon_\gamma(k) \rightarrow k^{\gamma-1}$ for small $k$ when $1
< \gamma <3$. When $\gamma > 3$, the low energy dispersion is dominated by
the $k^2$ term. When $\gamma \le 1$, $\varepsilon_\gamma(k)$ is not
well-defined in the thermodynamic limit; in order to have a well-defined
thermodynamic limit, the hopping energy $t$ must be properly scaled by
the system size in this case.

Next we will consider the thermodynamics of this model with different
$\gamma$ and demonstrate that for $\gamma < 2$, we indeed have a
finite temperature BEC. At low temperature, only the small
$k$ part of the spectrum is important. For $1 < \gamma < 3$ we consider free bosons
with a dispersion $\sigma k^{\gamma-1}$. Here $\sigma = -2 t \Gamma(1
- \gamma)$ is given in Eq. (\ref{Eq:vare_expansion}). The average
density of such system in the thermodynamic limit is given by:
\beq
\frac{ \la N \ra}{L} = \frac{1}{2 \pi}\int dk \ev{N_k} =
\frac{1}{2\pi} \frac{\sigma^{\frac{1}{1 - \gamma}} }{(\gamma - 1)}
\int_0^\infty d\varepsilon  \frac{\varepsilon^{\frac{1}{\gamma -1}
    -1}}{z^{-1}e^{\beta \varepsilon} - 1} = \frac{(\beta
  \sigma)^{\frac{1}{1-\gamma}}}{2 \pi (\gamma -1)}
\Gamma(\frac{1}{\gamma -1}) g_{\frac{1}{\gamma - 1}}(z),
\eeq
where $z = e^{\beta \mu}$, $\beta = \frac{1}{T}$ is the inverse
temperature, and $g_v(z) = \frac{1}{\Gamma(v)} \int_0^\infty dx
\frac{x^{v-1}}{z^{-1} e^x - 1}$ is the Bose-Einstein integral
function. To have a finite temperature BEC,
$\ev{N} / L = \ev{n}$ must remain finite when $z \rightarrow 1$, which
indicates $\frac{1}{\gamma - 1} > 1$, because for $v \le 1$,
$g_{v}(1)$ diverges.

\begin{figure}
  \includegraphics[width=8cm]{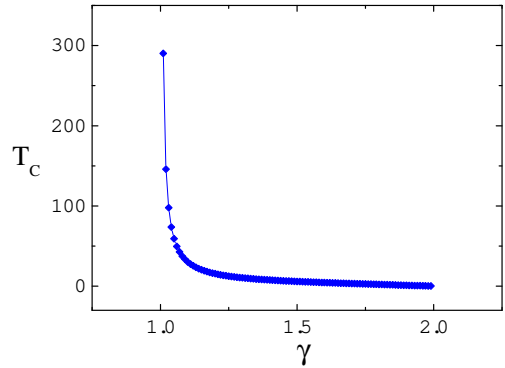}
  \caption{(Color online) Numerical calculation of $T_C$ (line + symbols) for the
    long-range hopping model in the thermodynamic limit. $T_C$ is
    measured in unit of the hopping energy $t$ which is set to 1. As one can
    see, $T_C$ grows monotonically from $0$ to $\infty$ as $\gamma$
    goes from $2$ to $1$. The divergent behavior of $T_C$ as $\gamma
    \rightarrow 1$ is a consequence of the divergent bandwidth.}
  \label{Fig:Tc}
\end{figure}

To have a better understanding of the thermodynamics of this
model, in Fig. (\ref{Fig:Tc}) we present a numerical calculation of
$T_C$. The exact average density of the system is:
\beq
\frac{\la N \ra}{L} = \frac{1}{2\pi} \int_{-\pi}^{\pi} dk \frac{1}{e^{\beta(\varepsilon_\gamma(k) - \mu)} - 1},
\eeq
where $\varepsilon_\gamma(k)$ is the eigenenergy function given in
Eq. (\ref{Eq:ek}). $T_C$ is computed by setting $\mu =
\varepsilon_\gamma(0)$ and then solving this equation
numerically. According to Fig. \ref{Fig:Tc}, $T_C$ grows
monotonically from $0$ to $\infty$ as $\gamma$ goes from $2$ to
$1$. The divergent behavior of $T_C$ as $\gamma \rightarrow 1$ is a
consequence of the divergent bandwidth in that limit.

According to our study of the infinite range hopping model, above
$T_c$ the mutual information should saturate as the system size
grows. Below $T_C$, the mutual information has a scaling
behavior $E_M \simeq \frac{1}{2} \ln L_A$; for $T = T_C$ $E_M
\simeq \frac{1}{4} \ln L_A$ for equally partitioned system in that model. We expect
the $\ln L_A$ scaling behavior both below $T_C$ and at $T_C$ to
persist in the long-range hopping model. As we shall see later, this
is indeed the case. However, the details of the scaling behavior
(i.e. the prefactor) can be different for different $\gamma$.
To study this scaling behavior, we fix the temperature and examine the
mutual information as a function of system size. This is desirable
because, if our conjecture according to the study of infinite-range
hopping model is correct, the mutual information will be
proportional to $\ln L_A$ when $T \le T_C$.

\begin{figure}
  \includegraphics[width=8cm]{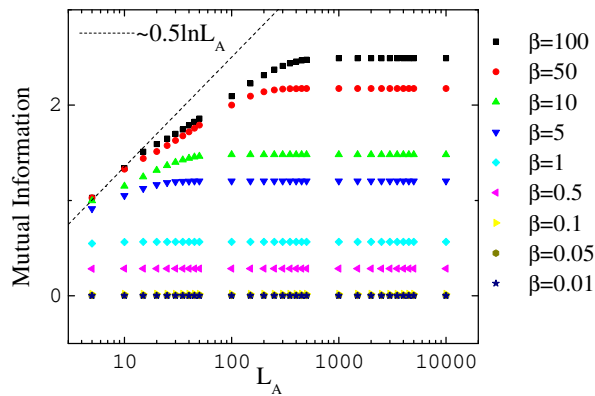}
  \caption{(Color online) Mutual information of the nearest neighbor hopping model
    plotted against subsystem size on a logarithmic scale. Average
    density is set to $\ev{n} = 1$, and the system is equally
    partitioned, $L = 2 L_A$. The black dash line is $E_M \sim
    \frac{1}{2}\ln L_A$. This line will be in other graphs for comparison as
    well. Clearly, the mutual information saturates when the system
    size grows large enough.}
  \label{Fig:fixT_k2}
\end{figure}

Before we demonstrate our results in the long-range hopping model,
first let us verify our analysis in the NN hopping
model in which no BEC would occur. This case actually corresponds to
$\gamma \rightarrow \infty$. The Hamiltonian is given by
restricting the hopping in Eq. (\ref{Eq:GeneralH}) to the nearest
neighbors only:
\beq
H_{n.n} = - t \sum_{\la ij \ra} \c_i^\dagger \c_j.
\eeq
Figure \ref{Fig:fixT_k2} is a linear-log plot of mutual information
against subsystem size at different temperatures. Throughout our study we shall consider equal
partition only, i.e., $L_A = L / 2$. The average density is also set to
$\ev{n} = 1$ here, so are the other results we will show
later. Clearly, at a fixed temperature, the mutual information
saturates as the system size grows. At low temperatures, small systems can be considered in the zero
temperature limit. This leads to the mutual information growing as
$\sim \frac{1}{2}\ln L_A$, until saturation kicks in.

\begin{figure}
  \includegraphics[width=8cm]{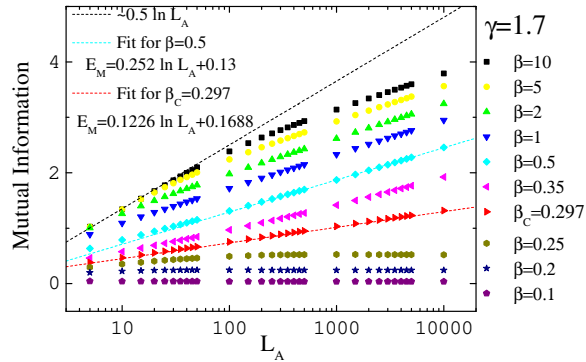}
  \caption{(Color online) Mutual information of the long-range hopping
    model with the parameter $\gamma = 1.7$ as a function of
    subsystem size on a logarithmic scale, at various (inverse)
    temperatures. The average boson density \ev{n} is set to 1, and
    the system is equally partitioned, $L_A = L /2$. The scaling behavior
    for inverse temperature $\beta = 0.5$ goes as $E_M = 0.2405 \ln L_A + 0.214$ (cyan
    dash line corresponding to the diamond data points). At the
    transition point, $\beta=\beta_C = 0.297$, the scaling law is fit
    to be $E_M = 0.1226 \ln L_A + 0.1688$ (red dash line corresponding
    to the right triangle data points).}
  \label{Fig:fixT_g1.7}
\end{figure}

Next we consider $ 1 < \gamma < 2$. Now the finite temperature transition
emerges, and the signature for the transition in mutual information -
the logarithmic scaling with (sub-)system size also emerges. In
Fig. \ref{Fig:fixT_g1.7}, we plot the mutual information (scatters) for $\gamma
= 1.7$ at different temperatures. The best fit (cyan dash
line corresponding to the diamond data points) for $\beta = 0.5 > \beta_C$
gives $E_M = 0.2405 \ln L_A + 0.214$. At $\beta_C = 0.297$, the
scaling behavior is fit (magenta dash line) as $E_M = 0.1226 \ln
L_A + 0.1688$. Both behaviors agrees qualitatively with what has been
suggested by our analytic study of the infinite-range hopping
model. When the temperature is well below $T_C$, we have the
logarithmic scaling behavior: a set of parallel linear lines on this
logarithm scale plot for different temperatures. However, the
prefactor is significantly different from that of the infinite range
hopping model which is $\frac{1}{2}$. In fact, by calculating mutual
information of different $\gamma$'s, we find that the prefactor varies as $\gamma$ changes.
For very low temperature, small systems again effectively fall into the zero
temperature region, and the mutual information restores to the $\sim
\frac{1}{2} \ln L_A$ behavior as in zero temperature. But when the
system size becomes large it crosses back to the finite temperature
scaling behavior again. This is evident for $\beta > 1$ in the
figure. For temperature close to but still below $T_C$, small systems
behave differently: the mutual information scales more like the line
for $\beta = \beta_C$, and it bends up as the system size increases and finally
crosses back to its genuine behavior below $T_C$. For temperature
close to $T_C$ but now above, small systems behave the other way: the
mutual information bends downwards and saturates at large system
size. $\beta_c$ serves as a very distinctive boundary between the
above two different bending behaviors. The latter two bending features
are also present in Fig. \ref{Fig:mi_inf}, our numerical verification of the infinite-range
hopping model. But the first feature at very low temperature is
missing in Fig. \ref{Fig:mi_inf} since in that case the mutual
information scales the same way as the entanglement entropy.

\begin{figure}
  \includegraphics[width=8cm]{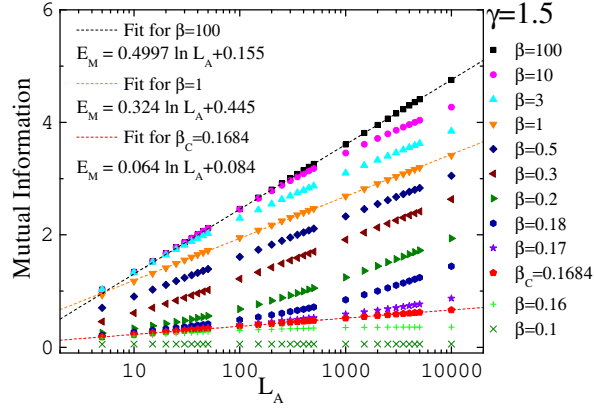}
  \caption{(Color online) Mutual information of the long-range hopping
    model with the parameter $\gamma = 1.5$ as a function of
    subsystem size on a logarithmic scale, at various (inverse)
    temperatures. The average boson density \ev{n} is set to 1, and
    the system is equally partitioned, $L_A = L /2$. The mutual
    information for $\beta = 1$ is fit to scale as $E_M \simeq 0.324
    \ln L + 0.445$ (orange dash line). At $\beta_C = 0.16843$, we observer a
    weaker scaling behavior which is fit to be $E_M
    = 0.064 \ln L_A + 0.084$ (red dash line).}
  \label{Fig:fixT_g1.5}
\end{figure}

\begin{figure}
  \includegraphics[width=8cm]{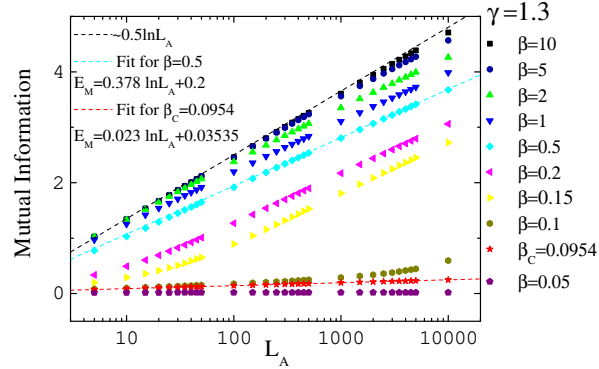}
  \caption{(Color online) Mutual information of the long-range hopping
    model with the parameter $\gamma = 1.3$ as a function of
    subsystem size on a logarithmic scale, at various (inverse)
    temperatures. The average boson density \ev{n} is set to 1, and
    the system is equally partitioned, $L_A = L /2$. The best
    fitting for $\beta = 0.5 > \beta_C$ (cyan dash line) gives $E_M = 0.378 \ln
    L_A + 0.2$. At $\beta_C = 0.0954$, the scaling behavior is fit
    (red dash line) as $E_M = 0.023 \ln L_A + 0.03535$.}
  \label{Fig:fixT_g1.3}
\end{figure}

Very similar behaviors were observed for the entire range $1 < \gamma <
2$. Representative results are presented in Figs. \ref{Fig:fixT_g1.5}
and \ref{Fig:fixT_g1.3} for $\gamma=1.5$ and $1.3$ respectively. We
thus conclude that for the entire range $1 < \gamma < 2$, mutual
information saturates for $T > T_C$, while it diverges logarithmically
with increasing subsystems size, for both $T < T_C$ and $T=T_C$. The
coefficients in front of the logarithms are $\gamma$-dependent.


\section{Summary and Concluding Remarks}

In this paper we have studied entanglement properties of free {\it
  non-relativistic} Bose gases. At zero temperature, all
particles fall into the ground state, and we find the entanglement entropy
diverges as the logarithm of the particle number in the
subsystem. At finite temperatures, we studied the
natural generalization of entanglement entropy - the mutual
information. We find the mutual information
has a similar divergence in the presence of a Bose-Einstein
condensate. When the system is above $T_C$ or does not have a
condensate, the mutual information saturates for large subsystem
size. It should be noted that for the special models we studied in
this paper there is no area-law contribution to the mutual
information, thus the contribution from the condensate, when present,
dominates the mutual information. In more generic models in two- or
three-dimensions where an area-law contribution is present, we expect
such logarithmic divergent contribution from the condensate to be
present as a sub-leading term in the subsystem-size dependence of the
entanglement entropy and mutual information.

Physically it is easy to understand why the condensate makes such an
important contribution to entanglement. First of all, BEC is
intrinsically a quantum process, just like entanglement reflects the
intrinsically quantum nature of the system. More specifically, when a
(macroscopically) large number of particles occupy the same state (at
$k=0$), they are necessarily delocalized throughout the sample, giving
rise to entanglement between blocks.

Just like in our previous work on a very different
system \cite{ding:052109}, our results here suggest that {\em conventional}
ordering, like BEC, makes a logarithmic
contribution to entanglement. One thus needs to take caution when using
entanglement as a diagnostic for exotic phases (such as topological
phases) or quantum criticality.

\section*{ACKNOWLEDGMENTS}

We thank Dr. Libby Heaney for a useful correspondence. This work was
supported by NSF Grant No. DMR-0704133. The authors thank the Kavli Institute for
Theoretical Physics (KITP) for the warm hospitality during the
completion of this work. The work at KITP was supported in part by
National Science Foundation Grant No. PHY-0551164.



\end{document}